\documentclass[reprint,amsmath,amssymb,aps,prb,twocolumn,superscriptaddress,showpacs,twoside,10pt,floatfix]{revtex4-1}
\usepackage[utf8x]{inputenc}
\usepackage[T1]{fontenc}
\usepackage{amssymb,amsmath}
\usepackage[]{graphicx} 
\usepackage{color}
\usepackage[colorlinks,bookmarks=false,citecolor=blue,linkcolor=blue,urlcolor=blue,backref=no,breaklinks=true]{hyperref}
\usepackage{bbm}
\usepackage{subfigure}

\newcommand{\et}{{\it et al. }}
\let\oldhat\hat
\renewcommand{\vec}[1]{\mathbf{#1}}
\renewcommand{\hat}[1]{\oldhat{\mathbf{#1}}}

\newcommand{\Erwa}[1]{\langle #1 \rangle}

\newcommand{\bi}{\begin{itemize}}
\newcommand{\ei}{\end{itemize}}

\newcommand{\ali}[1]{\begin{align}#1\end{align}}

\begin{document}
\title{Topological floating phase in a spatially anisotropic frustrated Ising model}

\author{Ansgar Kalz}
\affiliation{Institut für Theoretische Physik, Universität Göttingen, 37077
Göttingen, Germany} \email{kalz@theorie.physik.uni-goettingen.de}
\author{Gennady Y. Chitov}
\affiliation{Department of Physics, Laurentian University, Sudbury, ON, Canada}
\email{gchitov@laurentian.ca }

\date{\today}

\begin{abstract}
We present new results for the ordering process of a two-dimensional Ising
model with anisotropic frustrating next-nearest-neighbor interactions. 
We concentrate on a specific wide temperature and parameter region to confirm 
the existence of two particular phases of the model. The first phase is an incommensurate 
algebraically-ordered (floating) phase emerging at the transition from the paramagnetic
high-temperature phase. Then the model undergoes a transition to an antiferromagnetically ordered
second phase with diagonal ferromagnetic stripes (ordering wave vector $\vec q = (\pi/2, \pi/2)$).
We analyze the unconventional features appearing in several observables, e.g.,
energy, structure factors, and correlation functions by means of extensive Monte-Carlo 
simulations and examine carefully the influence of the lattice sizes. For the analytical study of the intermediate phase the Villain-Bak theory is adapted for the present model. Combining both the numerical and analytical work we present the quantitative phase diagram of the model, and, in particular, argue in favor of an intermediate topological floating phase.
\end{abstract}
\maketitle

\section{Introduction}
One of the most studied classical frustrated spin models is the
axial-next-nearest-neighbor Ising (ANNNI) model.\cite{P:fisher80, P:selke81,
P:selke88, P:villain81, P:bak82, P:rastelli10} The interplay of the frustrating
interactions of the model on a square lattice yields new physics in two
aspects: new phases emerge and the ordering processes are influenced by the
competition of the different states. In particular Fisher and
Selke\cite{P:fisher80} reported on the emergence of infinitely many
commensurate phases in a certain parameter region and Rastelli \et
\cite{P:rastelli10} demonstrated similar features of the ANNNI model very
nicely by applying Monte-Carlo (MC) simulations on finite lattices.

In the present work we examine another classical model which
has frustrating spin interactions through the diagonal next-nearest-neighbor
(nnn) bonds on the square lattice. This model was first studied by Fan and Wu.
\cite{P:fanwu69} They found, along the conventional ferromagnetic (FM) and
antiferromagnetic (AFM) phases, a new phase due to frustrating interactions
which they called superantiferromagnetic (SAF), and it is also more often
referred as columnar in current literature. This model was extensively studied
in the recent past \cite{B:liebmann86}, mainly due to the interest to the
predicted non-universality of the transition into the SAF phase. Even now
the model could harbor some surprises, as, e.g., the recently found phase 
transition change from the second to the first order at strong nnn coupling. 
\cite{P:lopez93, P:kalz11} 
We note however that the earlier work on this model was done almost exclusively for the
case of equal diagonal couplings. Motivated by some real material applications
\cite{P:chitovPRB04} one can generalize the above model for the anisotropic
case of the diagonal couplings of different strength or even of different
signs. It turns out that the anisotropic nn and nnn Ising model has a quite
rich phase diagram. \cite{P:chitov05} In particular, it possesses the
superferro-antiferromagnetic (SFAF) or $4\times4$ ground state phase with
ordering wave vector $\vec q_{44} = (\pi/2, \pi/2)$. After a $\frac{\pi}{4}$
rotation the ordering pattern of the $4\times4$ state becomes equivalent to the
antiphase of the ANNNI model, as one can see from Fig.~\ref{f:sketch}. From the
mean-field analysis and Kosterlitz-Thouless-type arguments
\cite{P:villain81,P:bak82} Chitov and Gros\cite{P:chitov05} also predicted the
incommensurate (floating) phase to be stable within a finite intermediate temperature
range above the low-temperature $4\times4$ phase. The same floating phase also
occurs in the 2D ANNNI model. \cite{P:villain81,P:bak82,P:selke88} It is
characterized by a lack of local order parameter and algebraic decay of 
correlation functions, modulated by plane wave oscillations with an
incommensurate wave vector $\vec q$ depending on couplings and temperature. Upon
cooling and reaching the boundary of the $4 \times 4$ phase, the wave vector
smoothly evolves towards its commensurate value. The incommensurate (symmetric) wave vector 
$q^{x,y} = \pi \kappa$ is determined by the density of
domain walls ($\kappa$) in the direction of the ferromagnetic diagonals (cf.
Fig.~\ref{f:sketch}). The transition from the floating to disordered phase
occurs via proliferation of dislocations (melting) of those walls.
\cite{P:villain81,P:bak82} It is analogous to the Kosterlitz-Thouless
transition in the classical XY model undergoing through decoupling of
topological defects (vortices).

The concept of incommensurate phases and phase transitions from incommensurate
to commensurate ordered states was already discussed in many versions of the
Ising model in two and three dimensions, see for example
Refs.~\onlinecite{P:bak80, P:ibaev11, P:sato99, P:zhang11}. We should stress
two qualitative differences between 2D and 3D cases. The 2D floating phase is
characterized by the algebraic order/decay of the correlation functions and
continuous variation of the ordering wave vector with temperature and/or
couplings. There exists a local order parameter in the 3D case. Also, as the ANNNI
model reveals, the variation of the ordering wave vector has discreteness
(which survives even the thermodynamic limit) referred to as the devil's
staircase.\cite{P:bak82,P:selke88} The number of steps depends on the internal
parameters, and thus the staircase reflects the intrinsic properties of the
system. Such devil's staircases have also been observed experimentally (see,
e.g., Ref.~\onlinecite{P:fischer78,P:ohwada01}). Of course a pure numerical
analysis of finite systems is often not enough to reveal the true discreteness
of devil's staircase, and some complimentary analytical work is highly
desirable. In this context it is worth noting that the 3D generalization of the
present model has quite interesting properties. In particular, competing
interactions between stacked Ising planes can results in the devil's staircase
with respect to the ordering wave vector in the stacking
direction,\cite{P:chitovCM04} observed in the experimental work of Ohwada
\textit{et al}.\cite{P:ohwada01}

Here we will present Monte-Carlo (MC) results of the spatially anisotropic
$J_1$-$J_2$-Ising model. In particular energies, specific heats, correlation
functions and their Fourier transform -- the structure factor will be
discussed. Strong numerical evidence for a floating phase within a finite temperature
region is given which has not been observed before. We show that the Villain-Bak 
theory \cite{P:villain81} can be straightforwardly extended for the present Ising model, 
giving predictions consistent with the MC results.

The manuscript is structured as follows: In Sec.~\ref{s:model} the model and
its known properties are introduced before we present in Sec.~\ref{s:mcr} the
MC and analytical results which focus mainly on the description and
characterization of an intermediate floating phase. In a concluding
Sec.~\ref{s:dis} we discuss our findings in the context of frustrated Ising
models.

\section{Model}\label{s:model}
The model is given by summing over all interactions between nearest neighbors
(nn) and next-nearest neighbors (nnn) of Ising-spin variables $S_i = \pm1$ on a
two-dimensional square lattice ($N = L \times L$, periodic boundary
conditions):
\begin{align}
H&= J_1 \sum_{\text{nn}} S_i\,S_j + J^a_2 \sum_{\text{nnn}^x}  S_i\,S_j  +
J^b_2 \sum_{\text{nnn}^y}  S_i\,S_j   \label{e:hamil}\,.
\end{align}
The interaction $J_1$ for nearest neighbors can be chosen negative or positive
which favors a ferromagnetic or antiferromagnetic Néel state as configuration
of total minimal energy, i.e., as a ground state. The $J_2$ are chosen to be of
opposite sign $sgn(J^a_2)=-sgn(J^b_2)$ for the two perpendicular directions. In
the following numerical analysis we will set always $J^a_2 = -J^b_2$ and will
choose $|J_2|$ as energy unit, i.e., temperatures and the nn coupling are given
mostly in units of $|J_2|$.

\subsection*{Ground States}\label{ss:aniso_states}

\begin{figure}[!t]
\begin{center}
\includegraphics[width=0.54\linewidth]{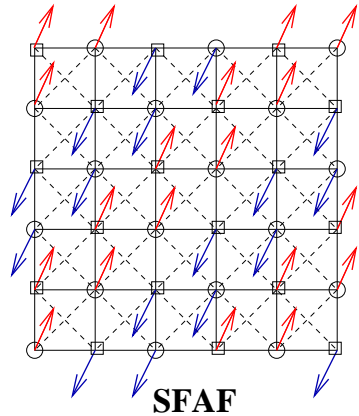}
\end{center}
\caption{\label{f:sketch} (Color online) The superferro-antiferromagnetic
(SFAF) or $4\times4$ ground state of the nn and nnn Ising model for $-|J_2| <
J_1 < |J_2|$.}
\end{figure}

For the given parameter set we expect the system to order in three different
ground states depending on the strength of the nn coupling
$J_1$.\cite{P:chitov05} For large antiferromagnetic coupling $J_1 > |J_2|$ the
Néel ordered (AFM) state is the ground state with a total energy of
$E_{\text{AFM}} = -2\,N\,J_1$ with all nnn bonds parallel aligned and all nn
bonds antiparallel aligned. On the other side of the phase diagram a
ferromagnetically ordered (FM) ground state with $E_{\text{FM}} = 2\,N\,J_1$
for $J_1 < - |J_2|$ is realized. In the intermediate region the SFAF
order yields the lowest energy where all nnn bonds are satisfied energetically,
i.e., antiparallel aligned in one direction and parallel aligned in the
perpendicular direction, and nn bonds are parallel and anti-parallel aligned
alternating in both directions, thus, the total energy of the nn sum gives zero
(see also Fig.~\ref{f:sketch}) and the ground-state energy is given by
$E_{\text{SFAF}}=-2\,N\,|J_2|$. The degeneracy of these ground states is
twofold (FM, AFM) or fourfold (SFAF). At the transition points $J_1 = \pm |J_2|$
the ground state is degenerate of order $L$ since every state which is
constituted out of $4\times 4$ plaquettes with a total spin $S=0$ yields the
same energy. So, the ground states have long-ranged orders, except at the
points of quantum criticality.

\subsection*{Phase Diagram}\label{ss:aniso_phase}
\begin{figure}[!t]
\includegraphics[width=0.48\textwidth]{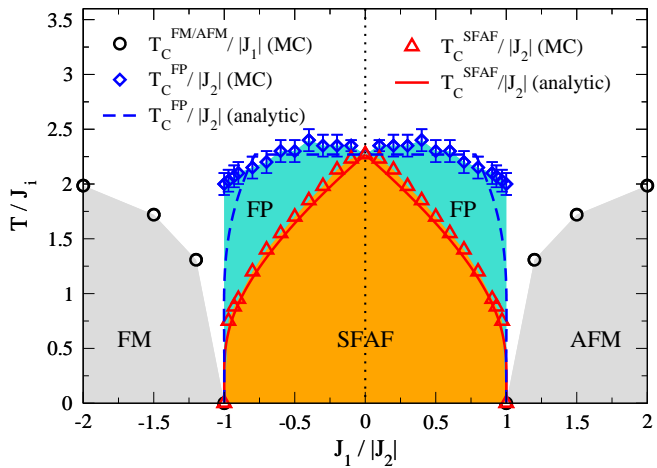}
\caption{\label{f:phase_aniso} (Color online) Phase diagram of the anisotropic
Ising model for varying nn interactions $J_1/|J_2|$. Three different ground
state configurations: Néel order (AFM) for $J_1 > |J_2|$,
superferro-antiferromagnetic order (SFAF) for $-|J_2| < J_1 < |J_2|$ and
ferromagnetic order (FM) for $J_1 < -|J_2|$ are separated by the quantum
critical points at $J_1 =\pm |J_2|$. The three phases (disordered, floating,
and $4\times4$) meet at the (exactly solvable) Lifshitz point at $J_1 =0$.
Critical temperatures are determined using Binder cumulants for $|J_1|
> |J_2|$ (energy scale is set to $|J_1|$ here) and estimated from energies,
specific heats and order-parameter behavior for $|J_1| < |J_2|$ (energy scale
is set to $|J_2|$). The transition temperatures in this regime strongly depend
on the system size -- here $L\leq 200$ (see text for more details). The red
solid line for $T_c^{SFAF}$ is obtained from Eq.~(\ref{TcF4}). 
The dashed blue line for $T_c^{FP}$ is given by the low-temperature 
Eq.~(\ref{kappaT}) and Eq.~(\ref{kappaC}) at 
$0.782(5) \lesssim J_1/|J_2| \leq 1$ and connects $T_c^{FP}$ at 
$J_1/|J_2| \approx 0.782(5)$ with the Lifshitz point as a guide to the eye. Note also the
agreement with a qualitative sketch in Ref.~\onlinecite{P:chitov05}.}
\end{figure}

The phase diagram for the model was introduced in Ref.~\onlinecite{P:chitov05}
for varying parameters $J_1$, $J_2^a$ and $J_2^b$. The result is a
three-dimensional qualitative phase diagram including ferromagnetic and various
antiferromagnetic phases (Fig.~2 of Ref.~\onlinecite{P:chitov05}). In the
present work the focus lies on a one-dimensional cut through this phase diagram
with a varying nn coupling $J_1$ and a fixed value $J_2^a = -J_2^b$. 
The finite-temperature phase diagram $T_c(J_1/|J_2|)$ was
qualitatively sketched in Fig.~6 of Ref.~\onlinecite{P:chitov05}. Here we
present a quantitative diagram in Fig.~\ref{f:phase_aniso}, obtained by the
direct MC simulations along with analytical work. Note that the transition
temperatures are invariant under the change of the sign $J_1 \rightarrow -
J_1$; this was double-checked in particular for $|J_1| < |J_2|$. For the
interactions $|J_1| > |J_2|$ the numerical critical temperatures were
determined using the Binder cumulants for different lattice
sizes.\cite{P:binder81L, P:binder81Z} From the intersection point of these
cumulants the critical point can be estimated.\cite{B:lanbin00, P:kalz08}

On the other hand for $|J_1| < |J_2|$, the extraction of critical temperatures
is more complicated since the order parameter for the $4\times4$ state and its
Binder cumulant show a strong finite-size dependence. Thus, for the estimation
of the critical temperatures in Fig.~\ref{f:phase_aniso} energies and specific
heats were also taken into account.

The nature of the phase transitions for $J_1 > |J_2|$ and $J_1 <- |J_2|$ is of
second order and belongs to the Ising universality class. However, according to
the prediction of Ref.~\onlinecite{P:chitov05}, the transition from the
high-temperature paramagnetic phase to the SFAF state is not direct but rather
involves an intermediate floating phase which will be the main topic of the
following sections.

\section{Monte-Carlo and Analytical Results}\label{s:mcr}
The MC simulations allow to calculate the energy, specific heat and various
order parameters for a wide temperature range and different parameters $J_1$
for finite lattice systems. Especially the analysis of the finite-size behavior
and the extraction of the values for the thermodynamic limit is crucial in the
investigation of incommensurate phases.
\begin{figure*}[!t]
\includegraphics[width=0.98\textwidth]{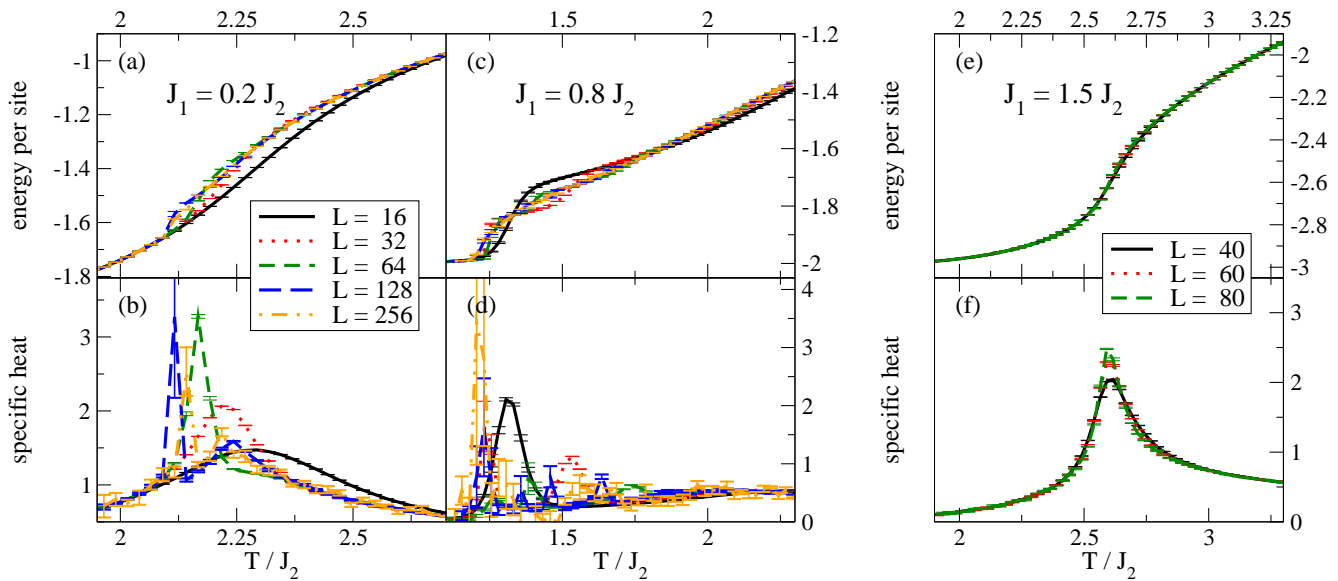}
\caption{\label{f:en_heat} (Color online) Energies and specific heats for some
values of nn coupling $J_1 > 0$. In Figs.~(a) and (c) multiple steps appear
which differ for different system sizes before reaching the ground-state energy
$E_{\text{SFAF}}= -2\,|J_2|$. As a comparison a converged energy development
for the phase transition to the AFM state is shown in (e) -- $E_{\text{AFM}}= -
2\, J_1$. The specific heats in (b) and (d) show multiple peaks which coincide
with the steps in the energies shown above whereas in (f) only a single peak is
observed which converges at the single transition temperature
$T_{\text{c,AFM}}$.}
\end{figure*}

For the simulations we used a Metropolis-single-spin
update\cite{P:metropolis53} with an additional exchange MC
step.\cite{P:hukushima96, P:hansmann97, P:katzgraber06} As a starting
configuration for several independent runs we selected the $4\times4$ state in
the appropriate parameter region for all temperatures and performed $10^6$
thermalization steps. This choice was reasoned in the large energy steps
between different states in the incommensurate region (see below) and already
proved itself very successful in a similar work on a frustrated Ising
model.\cite{P:rastelli10}

\subsection{Energy and Specific Heat}
The behavior of the energy already indicates that the ordering processes for
the two phase transitions -- to the AFM/FM phase and to the $4\times4$ state --
differ significantly. In Fig.~\ref{f:en_heat} the temperature dependence of the
energy and specific heat is shown for three cases: For $J_1 = 0.2\,|J_2|, \,
0.8\,|J_2|$ transitions to the $4\times4$ state are compared with an
Ising-phase transition at $J_1 = 1.5\,|J_2|$.
\begin{figure}[b!]
\includegraphics[width=0.48\textwidth]{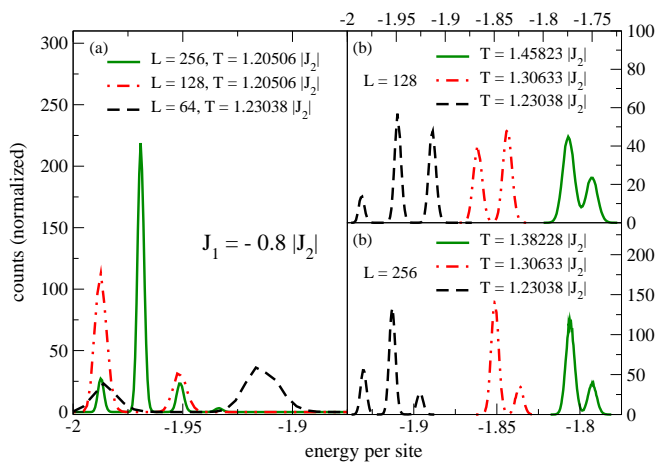}
\caption{\label{f:histo} (Color online) Energy histograms for some transitions
at $J_1 = 0.8\,|J_2|$: Double-peak features are very prominent for the
ground-state transition (a) and intermediate transitions at $L=128$ (b) and at
$L=256$ (c) as well. However, the energy gaps are not stable and seem to vanish
for all histograms in the thermodynamic limit. Note also the different energy
scales in panels (b) and (c).}
\end{figure}

\begin{figure*}[t!]
\includegraphics[width=0.98\textwidth]{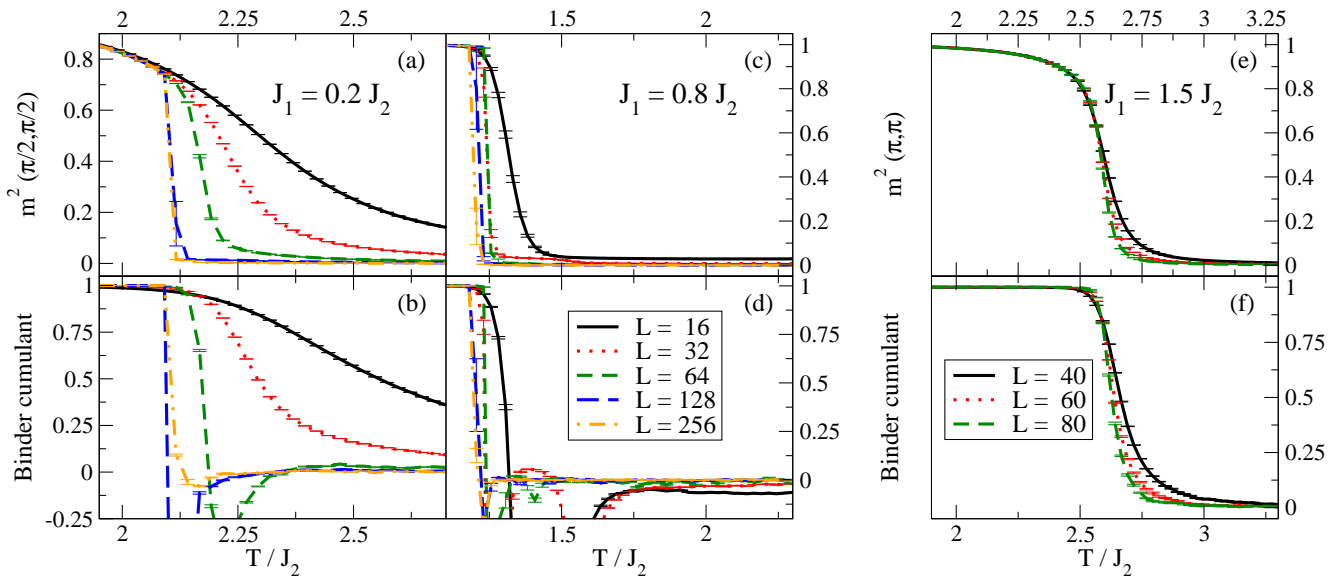}
\caption{\label{f:ord_bin} (Color online) Order parameters and Binder cumulants
for the $4\times4$ phase, i.e., moments of the structure factor
$S(\pi/2,\pi/2)$ for the same parameters as in Fig.~\ref{f:en_heat}. Although
in panels (a)-(d) a strong finite-size dependence is visible the convergence of
the observables is evident. However, an extraction of the exact transition
temperature to the ground state from the intersection of the different Binder
cumulants is only possible for the AFM case in panel (f).}
\end{figure*}

The multiple steps in the energies (Figs.~\ref{f:en_heat}(a) and (c)), which in
addition are shifted for different system sizes $L$, hint towards an unusual
ordering in the model that involves different size-dependent intermediate
states. This strong size dependence suggests already incommensurate ordering.

Even more prominent than the steps in the energy are the peaks in the
corresponding specific heat (Figs.~\ref{f:en_heat}(b) and (d)). They coincide
with the steps in the energy and prove the existence of several intermediate
states. The number of these states obviously depends on the system size; in
particular in Fig.~\ref{f:en_heat}(d) for each system size ($L =
16,\,32,\,64,\,128$) more peaks are distinguishable. However, another feature
is observed: The position of the last peak, i.e., for the lowest temperature
($T \approx 1.23\,|J_2|$), converges with increasing system size. A comparison
with the energy curve above (Fig.~\ref{f:en_heat}(b)) shows that below this
temperature the system is ordered in the $4\times4$ state according to the
energy value. Thus the final transition to the ground state is locked at a
finite temperature similarly to the conventional transitions into the AFM or FM
states (compare, e.g., Fig.~\ref{f:en_heat}(f)).

For a first characterization of the variety of transitions we recorded energy
histograms. In Fig.~\ref{f:histo} we show energy distributions at $J_1 =
0.8\,|J_2|$ for system sizes $L = 64$, $L = 128$ and $L=256$ at different
temperatures. As expected by the stepwise behavior of the energy, the
histograms show double peaks for selected temperatures (at the size-dependent
transitions temperatures between intermediate states and to the ground state).
The position of the ground-state transition is locked for the three histograms
shown in Fig.~\ref{f:histo}(a) but the peak-to-peak distance varies drastically
with the systems size: Essentially the gap between the prominent peaks in the
energy distribution shrinks by a factor two while the linear system size is doubled. The
same behavior can be extracted for the intermediate transitions. In
Figs.~\ref{f:histo}(b) and (c) histograms are shown which are recorded at
higher temperatures for (b) $L=128$ and (c) $L=256$. The multiple-peaked
features are again very prominent but the peak-to-peak distances shrink (note
the different energy scales for both figures) and the temperatures are shifted
as well.

Thus, according to the analysis of energy distributions all finite-size
transitions show strong first-order behavior.\cite{B:berg04} However, this
seems not to hold in the thermodynamic limit since the energy gaps tend towards
zero. Similar behaviour was observed in the 2D ANNNI model\cite{P:rastelli10} and 
for the isotropic version of the present model.\cite{P:kalz12}

\subsection{Order Parameters and Correlation Functions}
To gain further insight into the ordering process of the system it is useful to
define order parameters and analyze their behavior around the phase
transitions. The order parameters for the FM and AFM phases are readily defined
as magnetization and staggered magnetization which can also be expressed via
the spin-structure factors \ali{ S(\vec q) = \sum_{i,j} e^{i \vec q (\vec r_i -
\vec r_j)} S_iS_j } with ordering wave vectors $\vec q_{\text{FM}} = (0,0)$ and
$\vec q_{\text{AFM}} = (\pi,\pi)$. The wave vector for the SFAF phase is given
by $\vec q_{44} = (\pi/2, \pi/2)$; the (square) unit cell has four lattice spacings
in each direction (see Fig.~\ref{f:sketch}). The square root
of the normalized structure factor at this wave vector yields a good order
parameter. The calculation of this order parameter can be implemented also
using a staggered magnetization \ali{
m_{4\times4,k} = &\sqrt{\frac{S(\pi/2, \pi/2)}{N}} = \frac{1}{N}\sum_i (-1)^{f_k(r_i^ x,r_i^ y)}S_i \label{eq:order_SO} \\
&f_0(r^x,r^y) = [(r^x+r^y)/2] \%\,2 \nonumber \\
\wedge ~ &f_1(r^x,r^y) = [(r^x+r^y + 1)/2] \% \,2\,. } The modulo operation
'$\%\,2$' yields values zero and one, and the two versions $f_{0,1}$ account
for the degeneracy of the ground state, i.e., a shift of all spins by one
lattice spacing. In addition both states can be flipped completely. In
Fig.~\ref{f:ord_bin} this order parameter and its Binder
cumulant\cite{P:binder81L, P:binder81Z} \ali{ U_B=\frac{3}{2}\left(1 -
\frac{\Erwa{m^4}}{3\Erwa{m^2}^2}\right) } are shown for increasing lattice
sizes and nn couplings $J_1 = 0.2\,|J_2|$ (panels a, b) and $J_1 = 0.8\,|J_2|$
(panels c, d). As a comparison in panels (e, f) we show the behavior of the AFM
order parameter and its Binder cumulant; from the intersection point the
transition temperature can be extracted. However, for $|J_1| < |J_2|$, i.e.,
for the transitions to the $4\times4$ state such an analysis is hampered by
strong finite-site effects which cause an unusual behavior at the transition:
The cumulants do not intersect in a single point and exhibit several dips for
intermediate temperatures. To properly understand the numerical results we must
recall that the paramagnetic and the $4\times4$ phases are separated by the
intermediate floating phase\cite{P:chitov05} which, in particular, does not
have a conventional local order parameter.
\begin{figure}[t!]
\includegraphics[width=0.48\textwidth]{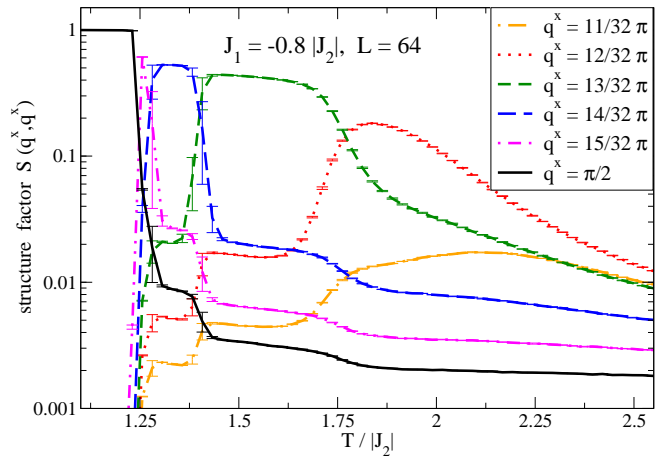}
\caption{\label{f:structs} (Color online) Structure factors $S(q^x,q^x)$ for a
series of momenta $0 < q^x \leq \pi/2$ vs. temperature. The transitions between
the intermediate states are clearly visible. Note also the stepwise behavior of
the signals which is only recognizable in the logarithmic scale.}
\end{figure}
%

\subsection{Floating Phase (analytical)}
The key to the analytical treatment of the floating phase comes from the
observation that the $4 \times 4$ phase in the ground state has exactly the
same pattern (2 rows of ``up" spins, then 2 rows of ``down" spins, or ($++--$)
for brevity) as the antiphase of the 2D ANNNI model (cf. Fig.~\ref{f:sketch})
when viewed along the ferromagnetic diagonals. So the Villain-Bak
theory,\cite{P:villain81} developed for that model, can be adapted here with
minor modifications. In the vicinity of the QCP ($J_1=|J_2|$) at low
temperature our model can be mapped onto an effective free-fermionic model
which accounts for the dynamics of the domain walls. In the floating phase
these walls are not straight anymore. The number density of domain walls
$\kappa$ is determined via minimization from the following equation 
valid at low temperature:
\begin{equation}
\label{kappaT}
    s(\kappa) = -2 \beta (J_2-J_1)\exp(2 \beta J_2)~,
\end{equation}
where
\begin{equation}
\label{sq}
    s(\kappa) \equiv \frac{1}{1-\kappa}  \cos \frac{\pi \kappa}{1-\kappa}
    - \frac{1}{\pi}  \sin  \frac{\pi \kappa}{1-\kappa}~,
\end{equation}
and from the physical meaning of $\kappa$, it is bound $0< \kappa <1/2$.
The correlation function in the direction perpendicular to ferromagnetic
diagonals decays algebraically in the floating phase
\begin{equation}
\label{SSfl}
    \langle S( \vec r' + \vec r)S( \vec r') \rangle \sim r^{-\eta} \cos (\vec q \cdot \vec r)
\end{equation}
with the power-law index
\begin{equation}
\label{eta}
    \eta = \frac12 (1- \kappa)^2~,
\end{equation}
and the wave vector of oscillations $q = \pi \kappa$ determined by the density
of walls $\kappa$. This floating phase is bound by the low-temperature SFAF
phase and the disordered (paramagnetic) phase at higher temperatures. The
critical temperature of the phase transition from the floating into the
$4\times4$ phase is given by the M\"{u}ller-Hartmann--Zittartz method in the
framework of the free fermionic approximation:\cite{P:villain81,P:muller77}
\begin{equation}
\label{TcF4}
    \sinh 2\beta J_2 \cdot \sinh 2 \beta (J_2-J_1)=1~,
\end{equation}
written here for $J_1,J_2>0$. As one can see from Fig.~\ref{f:phase_aniso},
this equation agrees nicely with the numerical MC results, reproducing even the
exactly known Ising value of $T_c$ at $J_1=0$ (the Lifshitz critical point).
Qualitatively, the phase transition from the floating into the $4\times 4$
phase can be understood as freezing of the domain walls into the ($++--$)
structure.

\begin{figure}[t!]
\includegraphics[width=0.48\textwidth]{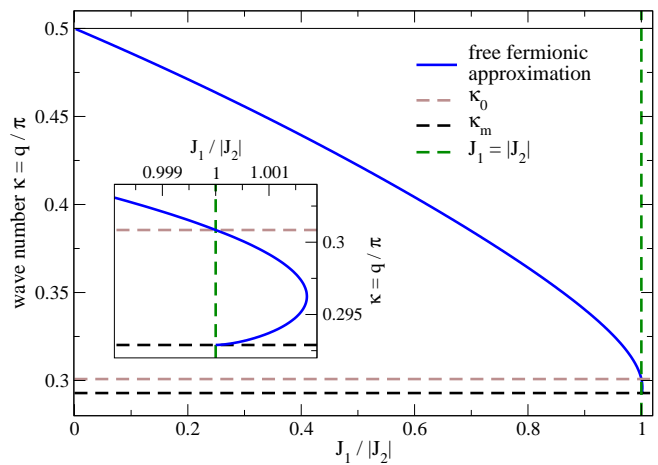}
\caption{\label{f:kappaC} (Color online) The critical domain wall density $\kappa_c=q_c/\pi$
($q_c$ is the critical ordering wave vector) as a function of couplings.
For each given coupling ratio the wave vector grows from $q=q_c$ at the critical temperature of the transition from the disordered to the floating phase $T_c^{FP}$ up to $q= \pi/2$ at the critical temperature $T_c^{SFAF}$ of the transition to the $4 \times4$ phase. The lower bound of $\kappa$, $\kappa_m=1-1/\sqrt{2}=0.2929$ and the reentrance boundary (see the text) $\kappa_0=0.3008$ are also shown.}
\end{figure}

On the other part of the diagram, with the increase of temperature the floating phase becomes
unstable and undergoes a phase transition into the disordered phase when the
dislocations start to proliferate in the network of the domain walls. In this
sense the phase transition from the (gapless) floating to the (gapped) disordered phase is
topological, not accompanied by the appearance/disappearance of any local order
parameter. It is analogous to the vortex unbinding transition in the classical
2D XY model, and the results within the approach due to Villain and Bak can be  
traced to their counterparts in the Kosterlitz-Thouless theory. 
The floating phase becomes unstable at the critical value of the walls density 
$\kappa_c$ (or at the wave vector $q_c= \pi \kappa_c$). 
It is determined from the following ansatz:
\begin{equation}
\label{kappaC}
  \kappa= 1- \frac{1}{\sqrt{2}[1+\exp(-4 \beta J_1)]^{1/2}}~.
\end{equation}
The critical temperature $T_c$ and density $\kappa_c$ are found from solution of the system of equations
(\ref{kappaT}) and (\ref{kappaC}). The result of the numerical solution of these equations 
for $\kappa_c$ (and thus for the critical wave vector $q_c=\pi \kappa_c)$ is given in  
Fig.~\ref{f:kappaC}.
$T_c^{FP}$ of the transition from the floating to the disordered phase is shown in Fig.~\ref{f:phase_aniso} by a dashed blue line. The low-temperature equation (\ref{kappaT}) does not work well at $J \ll |J_2|$, since the curve for $T_c$ ``overshoots" the exactly solvable Lifshitz point $J_1=0$. Contrary to Eq.~(\ref{TcF4}), Eq.~(\ref{kappaT}) does not cross over smoothly to the exact result at $J_1=0$ (note that $s(1/2)=-2$). This indicates the need for a better theory, taking into account, e.g., fermionic interactions at arbitrary temperature, which we relegate for future work.

In the vicinity of the QCP the system of equations (\ref{kappaT}) and (\ref{kappaC}) can be solved analytically. An interesting feature of the equations is a small reentrance effect, when the floating phase enters slightly into the FM and AFM domains $|J_1|>|J_2|$. The reentrance boundary is given by
$\kappa_c=\kappa_0$, where $\kappa_0 \approx 0.3008$ is the root of $s(\kappa_0)=0$. The critical temperature at the point of reentrance can be evaluated as:
\begin{equation}
\label{TcReen}
  \frac{T_c}{J_2} \bigg|_{|J_1|=|J_2|} \approx - 4/ \ln \big(2^{\frac32} (\kappa_0-\kappa_m) \big)
  \approx 1.05~,
\end{equation}
where $\kappa_m=1-1/\sqrt{2}=0.2929$ is the lower bound of $\kappa$. The bound follows from the stability condition for the floating (Kosterlitz-Thouless) phase:
\begin{equation}
\label{KTstab}
 \eta < 1/4 ~~\Longleftrightarrow~~\kappa > \kappa_m 
\end{equation}
One can also find that the critical temperature vanishes at the QCP as:
\begin{equation}
\label{TcQCP}
  \frac{T_c}{J_2} \approx 2/ \ln \Big( \frac{A}{|J_1/J_2|-1} \Big)~, 
\end{equation}
where 
\begin{equation}
\label{A}
  A \equiv  \frac12 s'(\kappa_0)(\kappa_0-\kappa_m)\approx 0.04~.
\end{equation}
At the QCP $\kappa \to \kappa_m$ and the correlation function index tends to 
the free fermionic (Ising) value $\eta \to 1/4$, while in the whole region of the floating phase $1/8 \leq \eta \leq 1/4$.
The effect of reentrance does not contradict the earlier prediction\cite{P:chitov05} that the floating
phase can be adjacent only to the SFAF phase (commensurability parameter $p=4$) 
of the present nn and nnn Ising model, since there are no commensurate-incommensurate 
transitions with small $p^2<8$, because then $\eta=2/p^2$ would violate the condition 
(\ref{KTstab}).\cite{P:bak82,P:selke88,B:liebmann86} Numerically, the reentrance is very small, 
and the reentrant regions do not overlap with the FM or AMF phases. It could well be an artefact 
of the low-temperature approximation (\ref{kappaT}), however, for a reliable test by MC simulations
parameter region is too small.

\subsection{Floating Phase (Monte-Carlo)}
To verify the predictions for the floating phase, we calculated the structure factors for symmetric momenta on the line $\vec q_{\text{FM}} = (0,0) \rightarrow \vec
q_{44} \rightarrow \vec q_{\text{AFM}} = (\pi,\pi)$. We observed finite
signals for some structure factors depending on the sign and value of $J_1$ and
more important also depending on the lattice size. Exemplary we show several
structure factors at $J_1 = -0.8\,|J_2|$ in Fig.~\ref{f:structs} in a
logarithmic scale for a broad temperature range. Due to the computational
effort here only $L=64$ is chosen, but nevertheless a cascade of signals can be
seen which starts at a critical momentum $q_c^x < \pi/2$ and ends up in the
ground-state order parameter with $q_{44}^x = \pi/2$. As we show above, the 
Villain-Bak theory for this model predicts that the intermediate floating phase 
is characterized by a single wave vector $\vec q$ while in the MC results 
for all momenta 
\ali{
\vec q_{\text{inc.}} &= (q^x_{\text{inc.}}, q^x_{\text{inc.}}) \quad \nonumber \\
&\text{and} \quad
\begin{cases}
q^x_c \leq q^x_{\text{inc.}} < \pi/2 & \text{for }J_1 < 0 \\
\pi/2 < q^x_{\text{inc.}} \leq q^x_c & \text{for }J_1 > 0 \label{e:qa_range}
\end{cases}\,.
} 
a non-zero contribution can be observed. The left-hand side of
Fig.~\ref{f:ani_corr} shows the structure factors for several $\vec q$'s also
in a color-coded plot for $L=32$ at $J_1 = -0.5|J_2|$ (upper left panel) and
$L=64$ at $J_1 = +0.8|J_2|$ (lower left panel). The behavior described in
Eqn.~\eqref{e:qa_range} can be seen very nicely in these plots. 
In conclusion in the intermediate phase the structure factor shows finite signals for certain
momenta which depend (i) on the sign and magnitude of the nn interaction and
(ii) on the lattice size. This can be explained by the simple fact that for a
discrete lattice the number of moments $q_{\text{inc.}}$ between $q_c (J_1)$
and $q_{44}=\pi/2$ increases with the system size.
\begin{figure}[t!]
\includegraphics[width=0.48\textwidth]{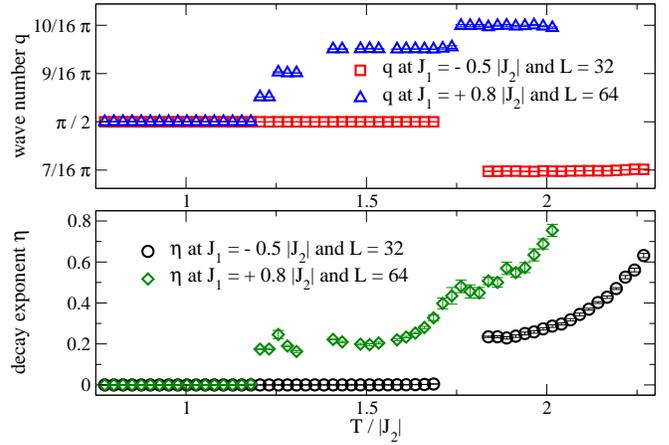}
\caption{\label{f:eta_q} (Color online) Results for the fit parameters of
Eqn.~\ref{SSfl} when applied to the finite systems for $L=32$ at
$J_1=-0.5\,|J_2|$ and for $L=64$ at $J_1=+0.8\,|J_2|$. In the upper panel the
development of the wave number is shown which reflects the behavior of the
structure factors given in the left-hand side of Fig.~\ref{f:ani_corr}. In the
lower panel the algebraic decay exponent is shown: For the intermediate
phase a saturation at different levels can be observed before it holds $\eta =
0$ in the ground state. (Values in the vicinity of transitions are left out
since there the fitting fails.)}
\end{figure}
Furthermore in Fig.~\ref{f:structs}, the stepwise behavior of the structure
factor at the transitions between different momenta is very prominent. However,
this feature is also due to the discrete spectrum of the momenta on a finite
lattice. As already observed in the energy histogram finite-size effects play a
crucial role in the intermediate phase. In the thermodynamic limit the temperature dependent momentum $\vec q(T)$ locks smoothly into the ground-state value $\vec q_{44}$.
\begin{figure*}[t!]
\subfigure[\label{f:ani_corr32} $L=32$;~$J_1=-0.5\,|J_2|$]
{\includegraphics[width=0.4\textwidth]{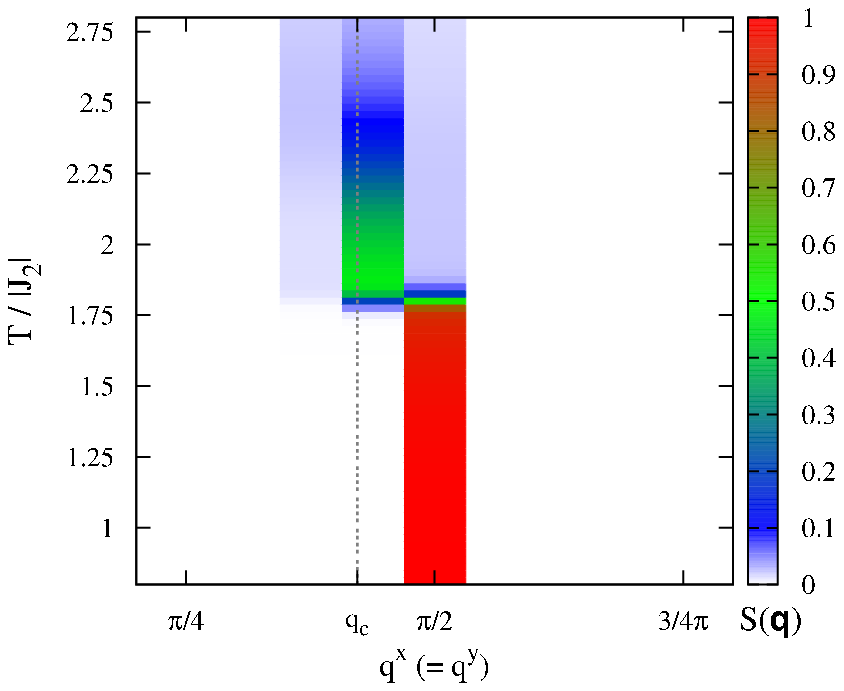}\quad\quad\quad\quad\quad
\includegraphics[width=0.4\textwidth]{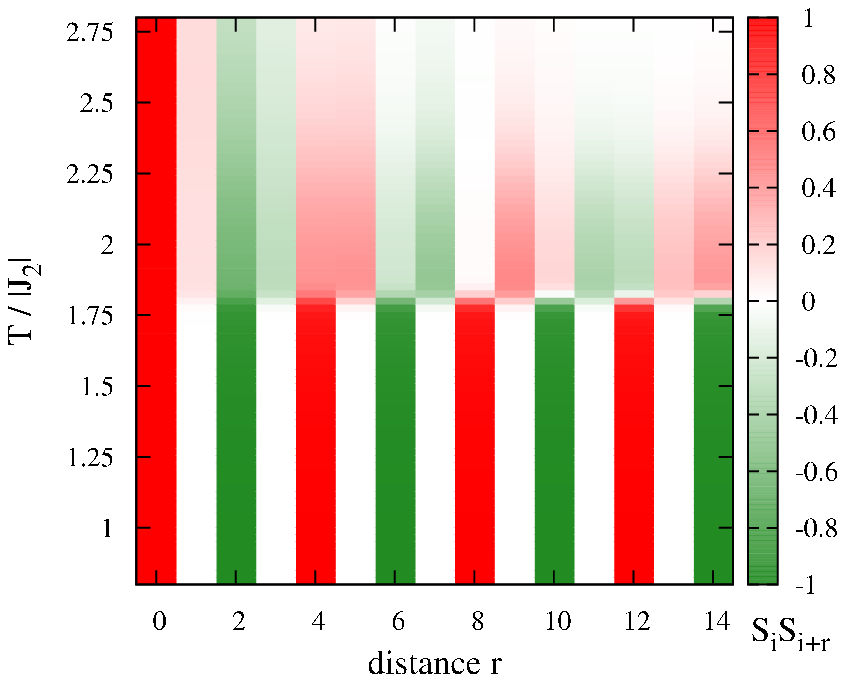}}\\
\subfigure[\label{f:ani_corr64} $L=64$;~$J_1=+0.8\,|J_2|$]
{\includegraphics[width=0.4\textwidth]{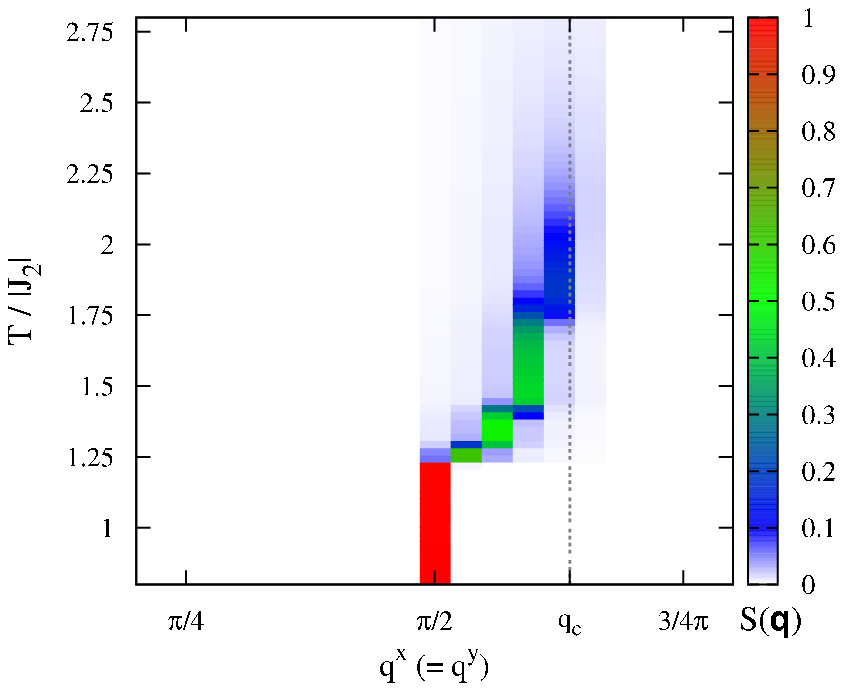}\quad\quad\quad\quad\quad
\includegraphics[width=0.4\textwidth]{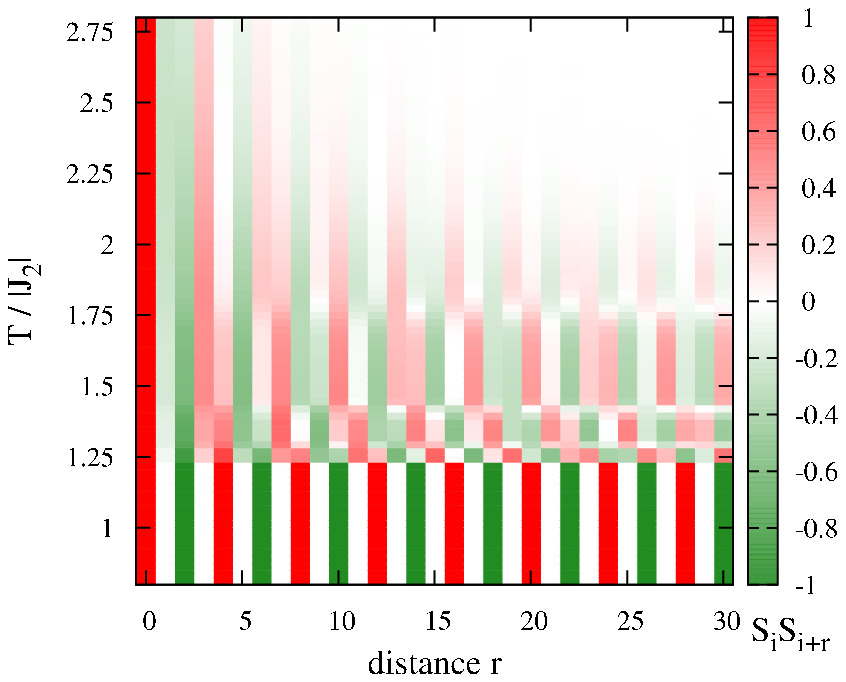}}
\caption{\label{f:ani_corr} (Color online) Structure factors (left) and
correlation functions (right) calculated from two different systems shown in
color code. The stepwise behavior of the structure factors coincides with the
distinguished temperature regions in the correlation functions, and the
features clearly depend on $J_1$ and $L$. The critical momentum $q_c$ is 
determined numerically as described above and shown in Fig.\,\ref{f:kappaC}.}
\end{figure*}

The size dependence of the signals in the structure factors  show already strong
indications for unconventional ordering in the system before the actual ground
state is reached. A further analysis relies on the evaluation of correlation
functions. On the right-hand side of Fig.~\ref{f:ani_corr} we present correlation
functions calculated in one row or column of the lattice for $L=32$ at $J_1 =
-0.5|J_2|$ (upper right panel) and $L=64$ at $J_1 = +0.8|J_2|$ (lower right
panel). These correlation functions clearly show \emph{non-fitting} oscillatory
behavior in the region of higher temperatures before the system orders in the
ground state with its four-site period. An analysis of these oscillations by
fitting Eqn.~\eqref{SSfl} at each temperature step yields a good agreement
in a wide temperature range. The resulting decay exponent $\eta$ and the wave
number $q$ can be extracted. Since the correlations are calculated in one
dimension only, the wave vector is reduced to one component here. Furthermore,
this wave number obviously coincides with the symmetric entries of the wave
vector of the corresponding structure factor which are shown on the left-hand
side of Fig.~\ref{f:ani_corr}. The results for $q$ (upper panel) and $\eta$
(lower panel) are given in Fig.~\ref{f:eta_q}.

In the plot fitting parameters in the vicinity of transitions and for high
temperatures are left out since at these points the fit fails and does not
yield meaningful results. The upper panel (wave numbers) only reproduces the
results from the structure factors but the lower panels shows that in the
intermediate phase the behavior of a floating phase is recovered. The fits of the decay
exponent give reasonable (taking into account the strong finite size effects) 
results in the floating phase. At low temperatures the exponent saturates at $\eta \sim 0.2$, 
while below $T_c$ into the $4 \times 4$ phase with the long-range order parameter 
it adopts the value $\eta = 0$.

\section{Conclusion}\label{s:dis}
An anisotropic version of the frustrated $J_1$-$J_2$ Ising model was
investigated using mainly MC simulations, supplemented by analytical treatments. 
In particular the phase transition into an
antiferromagnetic ground state constituted of two sublattices in collinear
order ($4\times 4$ phase) was analyzed. It was predicted earlier by Chitov and 
Gros\cite{P:chitov05} along with an unconventional floating phase which appears 
for intermediate temperatures before the ground state is reached. For analysis of the 
floating phase the Villain-Bak theory\cite{P:villain81,P:bak82}  with some small 
modifications was utilized for the present model. In the floating phase the theory 
predicts no local order parameter, algebraically decaying correlation functions 
with the power-law exponent $\eta$, modulated by a plane wave with a 
(single) incommensurate wave vector $q$ depending on couplings and temperature. 
The theory also allows to quantitatively describe the smooth evolution of the 
wave vector from its critical value $q_c$ at the critical temperature between 
the disordered and floating phases towards its commensurate value at the boundary 
of the $4 \times 4$ phase. The critical temperatures between these phases are also
evaluated. Qualitatively, the theory gives the picture of the 
transition from the floating to disordered phase via proliferation of dislocations
of the domain walls, which is analogous to the Kosterlitz-Thouless
transition of vortices in the classical XY model, while the transition into the 
commensurate phase occurs via freezing of the domain walls into the $4 \times 4$ 
structure.

The nature of the intermediate phase was also analyzed using correlation functions 
and the corresponding structure factors in the MC simulations.
It appears in the simulations of the finite-size system that the phase is not 
described by a single wave vector but rather a set of neighboring wave vectors. 
The transitions in finite lattices between states with different momenta are sharp, 
however, in the thermodynamic limit energy gaps seem to vanish. Thus the phase is best
described by a temperature dependent wave vector whose discrete spectrum is
smeared out in the thermodynamic limit. The momentum varies from a starting
vector $\vec q_c(J_1/ |J_2|)$ and locks smoothly into $(\pi/2,\pi/2)$.
The nature of this phase was further analyzed by fitting the correlation
functions by a combination of algebraic decay and oscillatory behavior. The
agreement in the intermediate phase is very good and we conclude from this
result that the state is best described by a floating phase consistently with 
the analytical predictions. The model's phase diagram in Fig.~\ref{f:phase_aniso}
summarizes most of our MC and analytical results.

As for the further work, we expect a very interesting behavior of the present model 
when the transverse field is included, especially near quantum criticality.\cite{P:chitov05}
Another interesting direction is the 3D generalization of the model, where the devil's 
staircase, similar to the one in the 3D ANNNI model,\cite{P:selke88,P:bak82}
is expected.\cite{P:chitovCM04} It appears that the devil's staricase in the 3D extension 
of the present model was already observed experimentally.\cite{P:ohwada01}

\begin{acknowledgments}
We thank the Deutsche Forschungsgemeimschaft for financial
support via the Collaborative Research Centre SFB 602 (TP A18, A.K.). The MC
simulations were performed on the parallel clusters of the Gesellschaft für
wissenschaftliche Datenverarbeitung Göttingen (GWDG) and of the North-German
Supercomputing Alliance (HLRN) and we thank them for technical support. G.Y.C.
acknowledges financial support from NSERC (Canada), Laurentian University
Research Fund (LURF), and the Ontario/Baden-W\"{u}rttemberg Faculty Exchange
Grant. G.Y.C. is grateful to the Institute for Theoretical Condensed Matter
Physics at Karlsruhe Institute for Technology, where a part of this work was
done, for hospitality. We acknowledge fruitful discussions with Sandro Wenzel
(EPF Lausanne) and Andreas Honecker (University of Göttingen).
\end{acknowledgments}

\bibliographystyle{apsrev4-1}

\bibliography{Literature_aniso}

\end{document}